\documentclass[preprint]{revtex4}
\usepackage{amsmath,amssymb,amsthm,graphicx,latexsym}
\newcommand{\bd}{\begin{document}}
\newcommand{\ed}{\end{document}}
\newcommand{\bc}{\begin{center}}
\newcommand{\ec}{\end{center}}
\newcommand{\bfr}{\begin{flushright}}
\newcommand{\efr}{\end{flushright}}
\newcommand{\lt}{\left}
\newcommand{\rt}{\right}
\newcommand{\vs}{\vspace}
\newcommand{\hs}{\hspace}
\newcommand{\beq}{\begin{equation}}
\newcommand{\eeq}{\end{equation}}
\newcommand{\lb}{\linebreak}
\newcommand{\pb}{\pagebreak}
\newcommand{\mb}{\makebox}
\newcommand{\fb}{\framebox}
\newcommand{\mc}{\multicolumn}
\newcommand{\ben}{\begin{enumerate}}
\newcommand{\een}{\end{enumerate}}
\newcommand{\bit}{\begin{itemize}}
\newcommand{\eit}{\end{itemize}}
\newcommand{\ovl}{\overline}
\newcommand{\un}{\underline}
\newcommand{\lefq}{\lefteqn}
\newcommand{\ba}{\begin{array}}
\newcommand{\ea}{\end{array}}
\newcommand{\beqa}{\begin{eqnarray}}
\newcommand{\eeqa}{\end{eqnarray}}
\newcommand{\beqas}{\begin{eqnarray*}}
\newcommand{\eeqas}{\end{eqnarray*}}
\newcommand{\bfg}{\begin{figure}}
\newcommand{\efg}{\end{figure}}
\newcommand{\bds}{\begin{displaymath}}
\newcommand{\eds}{\end{displaymath}}
\newcommand{\btb}{\begin{tabbing}}
\newcommand{\etb}{\end{tabbing}}
\newcommand{\para}{\parallel}
\newcommand{\pad}{\partial}
\newcommand{\nn}{\nonumber}
\newcommand{\la}{\leftarrow}
\newcommand{\ra}{\rightarrow}
\newcommand{\lgla}{\longleftarrow}
\newcommand{\lgra}{\longrightarrow}
\newcommand{\La}{\Leftarrow}\newcommand{\Ra}{\Rightarrow}
\newcommand{\Lra}{\Leftrightarrow}
\newcommand{\Lgla}{\Longleftarrow}
\newcommand{\Lgra}{\Longrightarrow}
\newcommand{\bm}{\boldmath}
\newcommand{\lan}{\langle}
\newcommand{\ran}{\rangle}
\renewcommand{\a}{\alpha}
\renewcommand{\b}{\beta}
\newcommand{\g}{\gamma}
\newcommand{\G}{\Gamma}
\renewcommand{\d}{\delta}
\newcommand{\eps}{\epsilon}
\newcommand{\s}{\sigma}
\newcommand{\D}{\Delta}
\newcommand{\vare}{\varepsilon}
\newcommand{\pr}{\prime}
\newcommand{\ro}{\rho}
\newcommand{\nab}{\nabla}
\newcommand{\m}{\mu}
\newcommand{\n}{\nu}
\newcommand{\Sg}{\Sigma}
\newcommand{\p}{\pi}
\newcommand{\R}{I\!\!R}
\newcommand{\om}{\omega}
\newcommand{\Om}{\Omega}
\newcommand{\ze}{\zeta}
\newcommand{\vart}{\vartheta}
\newcommand{\lam}{\lambda}
\newcommand{\tri}{\triangle}
\newcommand{\f}{\frac}
\newcommand{\iny}{\infty}
\newcommand{\pro}{\propto}
\newcommand{\np}{\newpage}

\begin{document}
\title{\large Pseudo Hermitian formulation of Black-Scholes equation}

\author{T. K. Jana}\email{tapasisi@gmail.com}\affiliation{Department of Mathematics, R.S. Mahavidyalaya, Ghatal 721212, India}, \author{P. Roy}\email{pinaki@isical.ac.in}
\affiliation{Physics and Applied Mathematics Unit, Indian Statistical Institute, Kolkata 700 108, India}
\vspace{2cm}

\begin{abstract}
We show that the non Hermitian Black-Scholes Hamiltonian and its various generalizations are $\eta$-pseudo Hermitian. The metric operator $\eta$ is explicitly constructed for this class of Hamitonians. It is also shown that the effective Black-Scholes Hamiltonian and its partner form a pseudo supersymmetric system. \\
\end{abstract}

%\pacs{PACS numbers:89.65.-s}
%\keywords{Keywords:Pseudo-Hermitian;Black-Scholes;Option Pricing}

\maketitle
\np
\section {Introduction}
During the past few years there has been a great interest in studying problems of finance using various tools of physics \cite{Mantegna}. In particular, different problems of finance have been studied from the point of view of quantum physics \cite{baaquie,Zhang,ilinski,schaden}. For example, various options have been modelled using quantum mechanical potentials  \cite{baaquie2}, option pricing with stochastic volatility has been studied using the path integral technique \cite{baaquie1}. Also, quantum mechanics has been used to analyze option pricing, stock market returns \cite{lin,cont,ali}, Black-Scholes (BS) equation \cite{haven,chou} etc. Supersymmetry formalism has been employed to obtain new solvable diffusion processes \cite{labo}.
\par
The BS equation (and its various generalizations) plays a dominant role in option pricing. The solutions of the BS equation may be found by mapping it into a Schr\"odinger like equation. Then various quantities like the pricing kernel or the option price can be obtained using the solutions of the Schr\"odinger like equation. It may be pointed out that from the point of view of quantum mechanics the BS Hamiltonian is non Hermitian.
On the other hand during the last decade non Hermitian quantum mechanics has been studied extensively. A feature of such systems is that Schr\"odinger equation with many non Hermitian potentials admit real eigenvalues \cite{bender}. Subsequently it was shown that this unusual feature may be attributed to $\cal{PT}$ symmetry \cite{bender} or more generally to $\eta$-pseudo Hermiticity \cite{mostafa}. Here our objective is to show that the BS Hamiltonian and its various generalizations are $\eta$-pseudo Hermitian and we shall determine the explicit form of the metric operator $\eta$ for each case. We shall also show that the effective BS Hamiltonian together with its partner Hamiltonian \cite{roy} form a pseudo supersymmetric system.

\section {$\eta$-pseudo Hermiticity of BS Hamiltonian}
The BS equation for option pricing with constant volatility is given by \cite{baaquie}
\beq
\frac{\partial C}{\partial t}=-\frac{1}{2}\sigma^2S^2\frac{\partial^2 C}{\partial S^2}-rS\frac{\partial C}{\partial S}+rC
\label{bs1}\eeq
where $C$, $S$, $\sigma$ and $r$ denotes the price of the option, the stock price, the volatility of the stock price and the risk-free spot interest rate respectively. Now under the transformations \cite{baaquie}
\beq
C(S,t)=e^{\epsilon t}\psi (S)~~~ \mbox{and}~~~~ S(x)=e^{x},~~-\infty< x < \infty
\label{tr1}\eeq
the BS equation (\ref{bs1}) becomes
\beq
\ba{l}
H_{BS}\psi=\epsilon \psi\\
H_{BS}=-\displaystyle\frac{\sigma^2}{2}\frac{d^2 }{dx^2}+\left(\frac{\sigma^2}{2}-r\right)\frac{d}{dx}+r
\label{bl2}\ea\eeq
where $H_{BS}$ is called the BS Hamiltonian. 
%\beq
%H\psi=[H_{BS}+V]\psi=\epsilon \psi
%\label{bsv}\eeq
It is well known that the BS Hamiltonian in Eq.(\ref{bl2}) can be brought to the Schr\"odinger form \cite{baaquie,Zhang}. To show this we use the similarity transformation
\beq
\rho H_{BS}\rho^{-1}=h_{BS} \label{tr2}
\eeq
where
\beq
\rho=exp\left[-\left(\frac{1}{2}-\frac{r}{\sigma^2}\right)x\right],~~~~h_{BS}=-\f{\sigma^2}{2}\frac{d^2}{dx^2} + \f{1}{2\sigma^2}\left(\f{\sigma^2}{2}+r\right)^2 \label{bl3}
\eeq\\
%and obtain from Eq. (\ref{blp})
%\beq
%h\phi=\lam \phi,~~~~h=-\frac{d^2}{dx^2} + \displaystyle \frac{2V}{\sigma^2},~~~~\lam=\frac{2\epsilon}{\sigma^2}-\left(\frac{r}{\sigma^2}+\frac{1}{2}\right)^2
%\label{bl3}\eeq
The Hamiltonian $h_{BS}$ in Eq.(\ref{bl3}) can be interpreted as a Schr\"odinger Hamiltonian of a particle of mass $\displaystyle\f{1}{\sigma^2}$ moving in a constant potential $\displaystyle\f{1}{2\sigma^2}\left(\f{\sigma^2}{2}+r\right)^2 $. It is important to note that the BS Hamiltonian in Eq.(\ref{bl2}) is non Hermitian while the Hamiltonian $h_{BS}$ in Eq.(\ref{bl3}) is Hermitian. We shall now show that the BS Hamiltonian $H_{BS}$ is in fact $\eta$-pseudo Hermitian.
\par

We recall that a Hamiltonian $H$ is said to be $\eta$-pseudo Hermitian if
\beq
H^\dagger =\eta H \eta^{-1}
\eeq
where $\eta$ is a Hermitian operator. It has been shown that eigenvalues of a $\eta$-pseudo Hermitian Hamiltonian are either completely real or occur in complex conjugate pairs \cite{mostafa}. In the context of financial modeling the BS equation usually has real eigenvalues and consequently it is of interest to examine the BS Hamiltonian from the point of view of pseudo Hermiticity.
\par
Let us now define the metric operator $\eta$ as
\beq
\eta =\rho^{2}= exp\left[-\left(1-\frac{2r}{\sigma^2}\right)x\right]\label{eta1}
\eeq
Then it can be verified that $\eta=\eta^\dag$ and
\beq
H_{BS}^\dag=\eta H_{BS} \eta^{-1}
\eeq
so that the BS Hamiltonian is $\eta$-pseudo Hermitian. Two important properties, namely, the completeness relation and the scalar product get modified for non Hermitian systems. In the present case they are given by
\beq
\sum_n \left|\psi_n\right>\left<\psi_n\right|\eta=1,~~~~\left<\psi_m|\psi_n\right>_\eta=\left<\psi_m|\eta\psi_n\right>=\int \eta(x) \psi_m^*(x)\psi_n(x)dx=\d_{mn}
\eeq

The above relations may be used to determine the pricing kernel. The pricing kernel $p(x,\tau,x^\prime)$ is defined as the conditional probability that the stock which has a value $e^x$ at time $t$ will have a value $e^{x^\prime}$ at time $T=t+\tau$. The pricing kernel for the BS Hamiltonian is then given by
\beq
p(x,\tau,x^\prime)=\left<x|e^{-\tau H_{BS}}|x^\prime\right>=\sum_{n}\eta(x^\prime) e^{-\tau\eps_n}\psi_n^*(x^\prime)\psi_n(x)
\eeq
Then the option price is given by
\beq
C(x,t)=\int \eta(x^\prime)  p(x,\tau,x^\prime) g(x^\prime)~dx^\prime
\eeq
where $g(x)$ is the pay off function.

\subsection{$\eta$-Pseudo Hermiticity of the Generalized BS Hamiltonian}

Sometimes the BS Hamiltonian can be generalized by including a security dependent potential $V(x)$. The resulting generalized Hamiltonian which satisfies the martingale condition is given by \cite{baaquie}
\beq
H=-\displaystyle\frac{\sigma^2}{2}\frac{d^2 }{dx^2}+\left(\frac{\sigma^2}{2}-V(x)\right)\frac{d}{dx}+V(x)\label{gbs}
\eeq
For an interpretation of the potential $V(x)$ from the point of view of finance we refer the reader to ref \cite{baaquie}. Now, the generalized BS Hamiltonian in (\ref{gbs}) is again non Hermitian. This can be seen from the fact that
\beq
H^\dag=-\displaystyle\frac{\sigma^2}{2}\frac{d^2 }{dx^2}-\left(\frac{\sigma^2}{2}-V(x)\right)\frac{d}{dx}+V^\prime(x)+V(x)\neq H
\eeq
The similarity transformation which transforms $H$ into the Schr\"odinger form is given by
\beq
\ba{l}
\rho=\displaystyle exp\left[\f{1}{\s^2}\int^x V(y)dy-\f{1}{2}x\right]\\\\
h=\displaystyle\rho H \rho^{-1}=-\f{\s^2}{2}\f{d^2}{dx^2}+\f{1}{2}V^\prime+\f{1}{2\s^2}\left(V+\f{1}{2}\s^2\right)^2
\ea
\eeq
To show the $\eta$-pseudo Hermiticity of the generalized BS Hamiltonian $H$ we define the metric operator $\eta$ as in the last section i.e,
\beq
\eta=\rho^2=\displaystyle exp\left[\f{2}{\s^2}\int^x V(y)dy-x\right]\label{eta2}
\eeq
Then $\eta=\eta^\dag$ and after some calculations it can be shown that 
\beq
\eta H \eta^{-1} = H^\dag
\eeq
so that the generalized BS Hamiltonian is $\eta$-pseudo Hermitian.

\subsection{$\eta$-pseudo Hermiticity of the effective BS Hamiltonians}

Path dependent options such as the Down-and-out barrier option, Out-and-out barrier option or the Double-knock-out barrier option can be analyzed by adding a potential term to the BS Hamiltonian (\ref{bl2}) and the effective Hamiltonian is given by 
\cite{baaquie,lin}
\beq
H_{eff}=H_{BS}+V(x)\label{eff}
\eeq
where $H_{BS}$ is given by (\ref{bl2}). Clearly for a real potential $V(x)$ we have
\beq
H_{eff}^\dag=H_{BS}^\dag+V(x)=-\displaystyle\frac{\sigma^2}{2}\frac{d^2 }{dx^2}-\left(\frac{\sigma^2}{2}-r\right)\frac{d}{dx}+r+V(x)\neq H_{eff}
\eeq
Since $V(x)$ is real, it is clear that in this case the metric operator is given by (\ref{eta1}) i.e,
\beq
\ba{l}
\eta = exp\left[-\left(1-\frac{2r}{\sigma^2}\right)x\right]\\\\
\eta H_{eff} \eta^{-1}=H^\dag
\ea
\eeq
In other words, the effective Hamiltonian $H_{eff}$ is $\eta$-pseudo Hermitian.

\section{Factorization of effective Hamiltonians}
It may be noted that factorization approach to effective Hamiltonians is often useful to find new exactly solvable processes \cite{labo,roy}. To use such a technique it is necessary to write the effective Hamiltonian as a combination of two operators $A$ and $B$ :
\beq
\ba{l}
H_{eff}=BA+\d\\
B=\displaystyle\f{\sigma}{\sqrt{2}}\left[-\f{d}{dx}+W(x)+(\f{1}{2}-\f{r}{\sigma^2})\right]\\
A=\displaystyle\f{\sigma}{\sqrt{2}}\left[\f{d}{dx}+W(x)-(\f{1}{2}-\f{r}{\sigma^2})\right]\\
\label{ab} \d=\displaystyle\f{1}{2\sigma^2}(\f{\sigma^2}{2}-r)^2+r \ea \eeq
where $V(x)=\displaystyle\frac{\sigma^2}{2}[W^2(x)-W^\prime(x)]$. Clearly $A$
and $B$ are not Hermitian conjugates of each other. However, it can
be shown that \beq B=\eta^{-1} A^\dag \eta\equiv A^{\#} \eeq so
that $A$ and $A^{\#}$ are $\eta$-pseudo adjoints of each other.
Consequently the effective BS Hamiltonian may be written as \beq
H_{eff}=A^{\#}A+\d \label{original} \eeq From Eq.(\ref{ab}) it is
possible to introduce the concept of the partner of the effective
Hamiltonian \cite{roy}. Reversing the order of the operators in
Eq.(\ref{original}) we find \beq H_{eff,P}=AA^{\#}+\d=H_{eff}+V_P
\label{partner} \eeq where
$V_P=\displaystyle\frac{\sigma^2}{2}[W^2(x)+W^\prime(x)]$. The Hamiltonian in
Eq.(\ref{partner}) may be called the partner of the effective
Hamiltonian. We shall now show that the
effective Hamiltonians $H_{eff}$ and $H_{eff,P}$ are related by pseudo
supersymmetry. To this end let us now consider an operator $Q$
defined by \beq Q=\left(\ba{cc} 0 & A\\0 & 0\ea\right) \eeq so
that its pseudo adjoint is given by \beq Q^{\#}=\eta^{-1} Q^\dag
\eta=\left(\ba{cc} 0 & 0\\A^{\#} & 0\ea\right) \eeq Then it can be
easily shown that \beq
\{Q,Q^{\#}\}={\cal{H}},~~~~[Q,{\cal{H}}]=[Q^{\#},{\cal{H}}]=0\label{pseudosusy}
\eeq where $\cal{H}$ is given by \beq
\cal{H}=\left(\ba{cccc}H_{eff,P}-\d & 0\\0 & H_{eff}-\d\ea\right) \eeq
Furthermore, it may be shown that \beq \eta
{\cal{H}}\eta^{-1}={\cal{H}}^\dag \eeq so that ${\cal{H}}$ is
$\eta$-pseudo Hermitian. The relations (\ref{pseudosusy})
constitute the $\eta$-pseudo supersymmetry algebra. Finally it may be noted that in case $\displaystyle r\rightarrow \f{\sigma^2}{2}$, the operators $A$ and $B$ defined in (\ref{ab}) become
\beq
\ba{l}
\displaystyle A=\f{\sigma}{\sqrt{2}}\left[\f{d}{dx}+W(x)\right]\\
\displaystyle B=\f{\sigma}{\sqrt{2}}\left[-\f{d}{dx}+W(x)\right]=A^\dag
\ea
\eeq
Also in this case $\eta=1$ and the pseudo supersymmetry algebra reduces to classical supersymmetry algebra. This is a consequence of the fact that in this limit the effective Hamiltonian (\ref{eff}) becomes Hermitian because of the absence of the first order derivative term. An explicit example can be found in ref \cite{roy}.
\section{Conclusions}
Here we have shown that the quantum BS Hamiltonian and it's various generalizations are $\eta$-pseudo Hermitian. The metric operator $\eta$ has also been found for each of these BS Hamiltonians. It has also been shown that the effective BS Hamiltonian and its partner Hamiltonian form a pseudo supersymmetric system. 

It may be recalled that here we have considered real potentials. However, in some situations (for example, to treat Asian options) it may be necessary to use complex potentials in the effective BS Hamiltonian \cite{coriano} and it would be of interest to examine ${\cal PT}$ symmetry or $\eta$-pseudo Hermiticity of such systems. Finally we feel it would also be of interest to investigate higher dimensional systems like the Merton-German Hamiltonians \cite{baaquie} from the point of view of $\eta$-pseudo Hermiticity.

%Let us now write the BS Hamiltonian (\ref{bl2}) in the form
%\beq
%H_{BS}=(q+i\b)^2+V(x)
%\eeq
%where
%\beq
%q=\f{\sigma}{\sqrt{2}}p,~~~~p=-i\f{d}{dx},~~~~\b=\f{1}{\sqrt{2}}\left(\f{\sigma}{2}-\f{r}{\sigma}\right)
%\eeq
%Then clearly $H_{BS}\neq H_{BS}^\dagger$ so that $H_{BS}$ is non Hermitian. As an ansatz for the metric $\eta$ we now consider \cite{zafar}
%\beq
%\eta=e^{f(x)},%
%\eeq
%$f(x)$ being a real function. Then we are to show that
%\beq
%H_{BS}^\dagger =(q-i\b)^2+V(x)=e^{f(x)}[(q+i\b)^2+V(x)] e^{-f(x)}=e^{f(x)} H_{BS} e^{-f(x)}\label{cond}
%\eeq
%After a starightforward calculation it can be shown that
%\beq
%q-i\b=e^{f(x)}(q+i\b)e^{-f(x)}=q+i\b+i\f{\sigma}{\sqrt{2}}f^\prime(x)\label{cond1}
%\eeq
%It is easy to see that if we choose $f(x)$ as
%\beq
%f(x)=-\f{2\sqrt{2}}{\sigma}\b x
%\eeq
%then (\ref{cond1}) will be satisfied. Thus it has been shown that $H_{BS}$ is $\eta$ - pseudo Hermitian with the metric given by
%\beq
%\eta = e^{-\f{2\sqrt{2}}{\sigma}\b x}
%\eeq

\ed
\begin{thebibliography}{99}

\bibitem{Mantegna} R.M. Mantegna and E. Stanley, Introduction to Econophysics, Cambridge University Press, (1999).
\bibitem{baaquie} B. E. Baaquie, Quantum Finance, Cambridge University Press, (2004).
\bibitem{Zhang} Y. Li, J. Zhang, Quantitative Finance {\bf 4}, (2004) 457.
\bibitem{ilinski} K. Ilinski, Physics of Finance, Wiley, 2001.
\bibitem{schaden} M. Schaden, Physica {\bf A316}, (2002) 511.
\bibitem{baaquie2} B.E. Baaquie et al, Physica {\bf A334}, (2004) 531.
\bibitem{baaquie1} B.E. Baaquie, J. Phys. I France {\bf 7}, (1997) 1733.
\bibitem{lin} V. Linetsky, Computational Economics {\bf 11}, (1998) 129; Mathematical Finance {\bf 3}, (1993) 349.
\bibitem{cont} M. Contreras et al, Physica {\bf A39}, (2010) 5447.
\bibitem{ali} A. Ataullah et al, Physica {\bf A388}, (2009) 455.
%\bibitem{field} B.E. Baaquie, Phys.Rev {\bf E64}, 016121 (2001); ibid {\bf E65}, 056122 (2002); ibid {\bf E75}, 016703 (2003); ibid {\bf E77}, 036106 (2008)\\
%B.E. Baaquie et al, Physica {\bf A334}, 531 (2004).
%\bibitem{baga} F. Bagarello, J. Phys. {\bf A 39}, 6823 (2006); Physica {\bf A 386}, 283 (2007); ibid {\bf 388}, 4397 (2009); Rep. on Math. Phys. {\bf 63}, 381 (2009).
\bibitem{haven} E.E. Haven, Physica {\bf A304}, (2002) 507; ibid {\bf A324}, (2003) 201; ibid {\bf A344}, (2004) 142.
\bibitem{chou} O.A. Choustova, Physica {\bf A374}, (2007) 304.
\bibitem{labo} P. Henry-Laboredere, Quantitative Finance {\bf 7}, (2007) 525.
%\bibitem{khare} F. Cooper, A. Khare and U. Sukhatme, Supersymmetry in Quantum Mechanics,
%World Scientific, Singapore, (2001).
%\bibitem{pursey} D.L. Pursey, Phys.Rev {\bf D36}, 1103 (1987).
%\bibitem{mielnik} B. Mielnik, J.Math.Phys {\bf 25}, 3387 (1984).
\bibitem{bender} C.M. Bender and S. Boettcher, Phys.Rev.Lett {\bf 80}, (1998) 5243.
\bibitem{mostafa} A. Mostafazadeh, J.Math.Phys {\bf A43}, (2002) 1418246 (10pp).
\bibitem{roy} T.K. Jana and P. Roy, Physica {\bf A90}, (2011) 2350.
\bibitem{coriano} B. E. Baaquie et al, Physica {\bf A334}, (2004) 531;
Also see pg 60 and pg 86 of ref \cite{baaquie}.
\end{thebibliography}
